\documentclass[12pt,letterpaper]{article}

\usepackage{amsmath,amssymb,array,calc,rotating,epsfig,psfrag, amscd}
\usepackage{color}
\usepackage[colorlinks=false,
    linktoc=all, %page 
    pdfstartview=FitV,
    bookmarksopen=true]{hyperref}
\usepackage[left=2cm,top=1cm,right=3cm,nohead]{geometry}
\usepackage[english]{babel}
\newcommand{\bea}{\begin{eqnarray}}
\newcommand{\eea}{\end{eqnarray}}
\newcommand{\be}{\begin{equation}}
\newcommand{\ee}{\end{equation}}
\newcommand{\barr}{\begin{array}}
\newcommand{\earr}{\end{array}}

\newcommand{\tphi}{\tilde{\phi}}
\newcommand{\txi}{\tilde{\xi}}

\newcommand{\non}{\nonumber}
\definecolor{cardinal}{rgb}{0.6,0,0}
\definecolor{darkgreen}{rgb}{0,0.5,0}
\definecolor{golden}{rgb}{0.92, 0.7, 0}
\definecolor{midnight}{rgb}{0, 0, 0.5}
\definecolor{darkblue}{rgb}{0.2, 0, 0.8}

\newcommand{\beq}{\begin{equation}\begin{aligned}}
\newcommand{\eeq}{\end{aligned}\end{equation}}
\newcommand{\nn}{\nonumber}

\numberwithin{equation}{section}

\renewcommand{\Re}{\text{Re} \;}
\renewcommand{\Im}{\text{Im} \;}

\begin{document}

\thispagestyle{empty}
 \begin{flushright}
 IPhT-t12/008
 \end{flushright}
\vspace{0.5cm}
\begin{center}
\baselineskip=13pt {\LARGE \bf{A comment on anti--brane singularities \\in
    warped throats\\}}
 \vskip1.5cm 
 \vskip0.5cm 
\renewcommand{\thefootnote}{\fnsymbol{footnote}}
 Stefano
 Massai\footnote{\href{mailto:stefano.massai@cea.fr}{\texttt{stefano.massai@cea.fr}}}\\ 
  \vskip0.5cm
 \textit{Institut de Physique Th\'eorique,\\
  CEA Saclay, CNRS URA 2306,\\
  F-91191 Gif-sur-Yvette, France}
\vskip0.5cm
\end{center}
\vskip1.5cm

\begin{abstract}
\noindent
We compute the imaginary self--dual (ISD) and imaginary anti--self--dual (IASD)
fluxes for the Klebanov--Strassler background perturbed by a stack of
$p$ anti--D3 branes. We show that, at linear order in $p$, they both
have a singularity in the near--brane region. 
While one can argue that the IASD flux may disappear at full
non--linear level, no such argument exists for the ISD mode. An
analogy with anti--D6 backreaction suggests that such singularity may
survive once full backreaction is taken into account and may be a
universal feature of anti--brane solutions.
\end{abstract}

\clearpage
%------%
%%%%%%%%%%%%%%%%%%%%%%%%%%%%%%%%%%%%%%%%%%%%%%%%%%%%%%

%%%%%%%%%%%%%%%%%%%%%%%%%%%
\section{Introduction}\label{secfirst}
%%%%%%%%%%%%%%%%%%%%%%%%%%%
\renewcommand{\thefootnote}{\arabic{footnote}}\setcounter{footnote}{0}

Recently, much attention has been devoted to the study of a
non--supersymmetric background (first
described by Kachru, Pearson and Verlinde~\cite{Kachru:2002gs})
obtained by  adding $p$ anti--D3 branes to the Klebanov--Strassler
solution~\cite{Klebanov:2000hb}. 
 This configuration is expected to be
metastable and to describe a metastable vacuum of the dual
$\mathcal{N}=1$ supersymmetric gauge theory.

To confirm this it is important to consider the effects of the
backreaction of the anti--branes and some preliminary steps were taken
in~\cite{DeWolfe:2004qx,DeWolfe:2008zy,McGuirk:2009xx}. The full
first--order  backreacted solution possibly describing these
anti--branes was obtained
in~\cite{Bena:2009xk,Bena:2011hz,Bena:2011wh}, and the 
parameters of this anti--brane solution were schematically described
in~\cite{Dymarsky:2011pm} and computed in~\cite{Bena:2011wh}. 
Analogous investigations have been performed
in similar models in
M--theory~\cite{Klebanov:2010qs,Bena:2010gs,Massai:2011vi} and type
IIA string theory~\cite{Giecold:2011gw}.

One of the key results of the analysis
of~\cite{Bena:2009xk,Bena:2011hz,Bena:2011wh} is that the first--order
backreacted anti--D3 brane solution must have a certain singularity in
the infrared. While there are no arguments that this singularity might
be physical, it has been suggested that it may be an artifact of
perturbation theory, and may go away once full backreaction is
taken into account~\cite{Dymarsky:2011pm,McGuirk:2009xx}. Indeed, as
we will review below, the imaginary anti--self--dual (IASD) flux couples to
the inverse warp factor, which is singular in perturbation theory but
is expected to be regular at the non--linear level.

In this paper we will show that besides a singular IASD flux, the
perturbative solution of~\cite{Bena:2009xk,Bena:2011hz,Bena:2011wh}
also contains a singular imaginary self--dual (ISD) flux for which
there is no argument that it will go away in the fully non--linear solution.

This is supported by the construction of the fully backreacted solution for anti--D6
branes in a flux background~\cite{Blaback:2010sj,Blaback:2011nz,Blaback:2011pn}, where a
singularity in the $H$-flux is unavoidable. This solution can be thought as a toy model
for ours: by T-dualizing it three times along the D6 worldvolume one
obtains a solution for anti--D3 branes in $\mathbb{R}^3 \times T^3$,
which is an increasingly better approximation of the KS near--tip
region as the $S^3$ radius grows or as the number of anti--D3 branes
becomes ever more smaller than the number of fractional branes.
 
Hence, our result suggests that at least the singular ISD flux, already visible
in the linearized solution, will persist in the fully backreacted
regime. It would be interesting to verify this by doing a fully
backreacted calculation. The problem would then be whether this
ISD singularity can be resolved, and it has been hinted~\cite{Dymarsky:2011pm} that this
might happen via brane polarization in the near--brane region~\cite{Myers:1999ps}, much like in
Polchinski--Strassler~\cite{Polchinski:2000uf}. At first glance this
seems unlikely, at least for smeared branes: the divergent
ISD flux we find does not seem to have the appropriate leg needed
for brane polarization.

%%%%%%%%%%%%%%%%%%%%%%%%%%%
\section{Anti--D3 branes on the deformed conifold}\label{KSback}
%%%%%%%%%%%%%%%%%%%%%%%%%%%

We will consider the setup in which a stack of $p$ anti--D3
branes has been introduced in the warped deformed conifold. The
dynamics of these anti--branes has been studied in the probe
approximation in~\cite{Kachru:2002gs}. Due to the warping and the five--form flux the
anti--branes are attracted to the tip of the geometry (namely, a
three--sphere $S^3$ of radius $\epsilon$, related to the confinement
scale of the dual gauge theory) and there they polarize (by
the Myers effect~\cite{Myers:1999ps}) into an $NS5$--brane wrapping a two--sphere $S^2
\subset S^3$. This state will decay to a supersymmetric state, but for
sufficiently small $p/M$ (where $M$ denotes the units of RR 3-form flux over the $S^3$) the probe analysis shows that this process  is non--perturbative and
the leading channel is via bubble nucleation, thus providing a
realization of a metastable supersymmetry--breaking state. From the
gauge theory point of view, this setup would describe a dynamical
supersymmetry breaking scenario in the Klebanov--Strassler
theory. A description of such a state from the gauge
theory side is lacking due to the strong coupling regime, thus
motivating the search for a gravity dual.

We now want to describe the effects of the backreaction of the stack
of anti--D3 branes on the ambient geometry. In order to make the calculation tractable, one can
make two simplifying assumptions: i) smearing the sources on the $S^3$
of the deformed conifold and ii) working in perturbation theory around
the supersymmetric background, at first--order in the parameter $p/M$, which we
suppose small in order to avoid perturbative decay to the
supersymmetric vacuum.

 In this approximation it is possible to find
an analytic solution in terms of integrals for the metric and flux modes that perturb the
Klebanov--Strassler background. This solution was obtained
in~\cite{Bena:2011hz}, building on the results
of~\cite{Bena:2009xk}. The anti--D3 solution was then obtained in~\cite{Bena:2011wh} by imposing appropriate boundary conditions on the space
of linearized deformations around the warped deformed conifold. Let us
briefly review the main conclusions of this analysis.\\

\noindent\textbf{IR behavior.} The boundary conditions that we should
impose in the near--brane region are those consistent with smeared
anti--branes at the tip: singular warp factor and five--form flux,
coming from the anti--D3 source with equal mass and charge, and regularity in all other modes. This
requirement fixes half of the sixteen integration constants of the
general linearized deformation around the conifold in terms of a
physical quantity: the number $p$ of anti--D3 branes at the tip. In particular,
this fixes the value of the mode (indicated by $X_1$
in~\cite{Bena:2011wh}) that gives rise to the force felt by a probe D3
brane in the backreacted geometry
\begin{equation}
F_{D3} = \frac{8\, 2^{2/3}\,\pi\,p}{h_0^2}\,  j'(\tau) \, ,
\end{equation}
where $j(\tau)$ is the Green's function at the linear order, defined
in~\eqref{Gintegral}. This agrees with the computation \`a la KKLMMT~\cite{Kachru:2003sx}
and is a nice check that the boundary conditions are the correct ones
to describe anti--branes. Once all those requirements are fulfilled,
one finds that remnant nonzero perturbations to three--form fluxes near
$\tau=0$ cause the energy density of such fluxes to diverge~\cite{McGuirk:2009xx,Bena:2009xk}. We stress
that this is purely an infrared phenomenon, in the sense that the
presence of the singularity is insensitive to the UV
boundary conditions. The scope
of this note is to dissect this singularity.

%%%%%%%%%%%%%%%%%%%%%%%%%%%
\section{Three-form flux singularities }\label{secfirst}
%%%%%%%%%%%%%%%%%%%%%%%%%%%

The Klebanov--Strassler background is parametrized by eight scalars
$\phi^a(\tau)$ which take values in a scalar manifold
$\mathcal{M}$ and which satisfy a first--order system of
ODE's~\cite{Klebanov:2000hb}. The space of linearized deformations
around this solution, which preserve its $SU(2)\times SU(2)\times
\mathbb{Z}_2$ symmetry, is described
by a system of eight second--order Euler--Lagrange equations for the perturbation modes
$\delta \phi^a(\tau)$. It is useful to use a Hamiltonian approach and
recast this system in terms of sixteen first--order equations for the
modes $\delta \phi^a(\tau)$ and their conjugates variables
$\xi^a(\tau)$. As a consequence of perturbing around a first--order
system, the equations of motion for the $\xi^a(\tau)$ modes
decouple~\cite{Borokhov:2002fm}. These modes provide a useful description of the
general solution since they parametrize supersymmetry breaking\footnote{While often stated in the
  literature, this is not quite true. The modes
  $\xi^a$ parametrize the breaking of the first--order description for
the zeroth--order solution, which for  the Klebanov--Strassler case is not equivalent to the 
conditions imposed by supersymmetry. In particular there exist
solutions to the first--order system with a non--vanishing
$(0,3)$--flux which break the supersymmetry~\cite{Kuperstein:2003yt,Halmagyi:2011yd}. As a result, if all the
$\xi^a$ are zero, the perturbation is not necessary
supersymmetric. The converse is however true, if any $\xi^a$ is different from zero, the perturbation breaks supersymmetry. In our case the only supersymmetry--breaking solution with $\xi^a = 0$ is
divergent in the UV.}. We will organize the metric and flux modes as
follows
\begin{equation}
\left\{ \Phi_{\pm}, \,\, G_{\pm}, \,\phi,
  \,g_{mn}\right\} \, ,
\end{equation}
and we will concentrate on the modes $\Phi_{\pm}$, $G_{\pm}$, defined
as
\begin{equation}
G_{\pm}  = \star_6 G_3 \pm i G_3\, ,\qquad \Phi_{\pm} = e^{ 4A} \pm \alpha \, ,\label{Gphidef}
\end{equation}
where $G_{\pm}$ are the ISD and IASD parts of the three--form flux, $\alpha$ is the RR 4--form and $e^{-4 A}$ is the warp factor
 (we refer to
appendix~\ref{notation} for the notations). The
dynamics of these modes is described by the equation of motion~\cite{Giddings:2001yu}
\begin{equation}\label{eomflux}
(d + i \frac{d\tau}{\Im \tau}\wedge \Re)(\Phi_{-} G_{+}+\Phi_{+}G_{-})
= 0 \, .
\end{equation}
In our Ansatz we have $\tau=e^{-\phi}$ since $C_{0} = 0$\footnote{The
  axion/dilaton $\tau$ in equation~\eqref{eomflux} should not be
  confused with the radial direction of the conifold.}.

\subsection{Linearized perturbations}\label{first}

We now want to compute the ISD and IASD flux once the backreaction of
the anti--D3 branes is taken into account. As explained in the previous
section, we linearize the problem by expanding in the parameter $\gamma
= p/M$:
\begin{align}
G_{\pm} & = G_{\pm}^0 + \, G_{\pm}^1(\gamma) + \mathcal{O}(\gamma^2) \, , \\
\Phi_{\pm} & = \Phi_{\pm}^0 + \, \Phi_{\pm}^1(\gamma) + \mathcal{O}(\gamma^2) \, .
\end{align}
For the Klebanov--Strassler background we have
\begin{equation}\label{KSminus}
\Phi^0_{-} = G^0_{-} = 0, 
\end{equation}
while 
\begin{align}
\Phi^0_{+} &= \frac{2}{h(\tau)} \, ,\label{KSphiplus}\\
G_{+}^0& = \left( f_0 - k_0 \right)
\left(g_1\wedge g_3\wedge g_5 + g_2\wedge g_4\wedge g_5 + i g_1\wedge g_3 \wedge
  g_6 + i g_2 \wedge g_4 \wedge g_6\right) \label{GPlusZero}\\
&\quad + 2i (2P-F_0) \,g_3\wedge g_4\wedge g_5 + 2i
F_0\, g_1\wedge g_2 \wedge g_5 \non\\
&\quad+ 2e^{2y_0} (2P-F_0)\,g_1\wedge g_2 \wedge g_6 + 2 e^{-2y_0} F_0\,
g_3\wedge g_4\wedge g_6 \, , \non
\end{align}
where the function $h(\tau)$ is the KS warp factor defined in~\eqref{hintegral}, while the other KS functions are given
in~\eqref{KSbackground}. At the linear order in $\gamma$, by using the expansions
of~\cite{Bena:2011wh} (to which we refer for details of the
solution) we find the flux modes:
\begin{align}
G_{-}^1 & = 2 e^{-4A_0} \Big[ \left( i g_1\wedge g_2 \wedge g_5
- e^{-2y_0} g_3 \wedge g_4 \wedge g_6 \right)\left(\txi_5 - \txi_6 \right) \label{Gmfirstorder}\\
&\qquad -\left( e^{2y_0} g_1\wedge g_2 \wedge g_6 - i g_3
  \wedge g_4 \wedge g_5 \right) \left(\txi_5 +\txi_6\right) \nn\\
&\qquad - \left( g_1 \wedge g_3 \wedge g_5 + g_2 \wedge g_4
  \wedge g_5 - i g_1 \wedge g_3 \wedge g_6 - i g_2 \wedge g_4 \wedge
  g_6\right) \, \txi_7\Big]\non \, ,
\end{align}
\begin{align}
G_{+}^1 & = e^{-4A_0}\Big[2ig_3\wedge g_4 \wedge g_5 \left(\txi_5 + \txi_6 -
  e^{4A_0}\tphi_7\right) + 2i g_1 \wedge g_2 \wedge g_5 \left(\txi_5 -
  \txi_6 + e^{4A_0}\tphi_7\right) \label{Gpfirstorder}\\
& \quad + 2 e^{2y_0} g_1 \wedge g_2 \wedge g_6 \left(\txi_5 + \txi_6
  + e^{4A_0} (4P\tphi_2 - 2F_0\tphi_2 - \tphi_7)\right)\nn \\
&\quad + 2e^{-2y_0}g_3\wedge g_4 \wedge g_6 \left(\txi_5 -\txi_6
  -e^{4A_0}(2F_0\tphi_2-\tphi_7)\right) \nn \\
& \quad- \left( g_1 \wedge g_3 \wedge g_5 + g_2 \wedge g_4
  \wedge g_5 + i g_1 \wedge g_3 \wedge g_6 + i g_2 \wedge g_4 \wedge
  g_6\right) \cdot \non \\
&\qquad \cdot\left(2 \txi_7 +
   e^{4A_0}(\tphi_6 - \tphi_5 + (f_0 - k_0) \tphi_8)\right)\Big] \, .\nn
\end{align}
Here the modes $\txi^a$ and $\tphi^a$ are respectively linear
combinations of the conjugate--momenta $\xi^a$ and the 
perturbations modes $\delta \phi^a$ (we refer to appendix~\ref{notation} for their definition).
By using the definition~\eqref{Gphidef}, we find the expressions for the
$\Phi_{\pm}$ modes at the linearized level in terms of the modes
$\txi^a$, $\tphi^a$
\begin{align}
\frac{ d\Phi_{-}^1 }{d\tau} &= \frac23 e^{-2x_0 (\tau)} \txi_1  \, , \\
\frac{ d\Phi_{+}^1}{d\tau} &=- \frac23 e^{-2x_0}\txi_1 + \frac{4\tphi_4
  h'(\tau) -4 h(\tau)\tphi_4'}{h(\tau)^2} \, .
\end{align}
The first equation can be integrated by using the equation of motion
for $\txi_5$~\eqref{xi5eom} and gives
\begin{equation}
\Phi_{-}^1 = -\frac{2}{P}\txi_5 +\text{const} =\frac{32}{3} X_1 j(\tau) + \text{const} \, ,\label{firstminus}\\
\end{equation}
where $P=M/4$ and  $j(\tau)$ is the Green's function defined
in~\eqref{Gintegral} and $X_1$ is proportional to the number of
anti--branes $p$
\begin{equation}\label{X1def}
X_1 = \frac{ 3\,\pi}{4 \, h_0^2}\,p \, ,
\end{equation}
where $h_0=18.2373\,P^2$. The equation for $\Phi_{+}^1$ can be easily
integrated to get
\begin{align}
\Phi_{+}^1 &=- \Big(\frac{32}{3} X_1 j(\tau) + \frac{4\,\tphi_4}{h(\tau)} \Big)+ \text{const}\, .\label{firstplus}
\end{align}
We note that $G_{-}^1$ and $\Phi_{-}^1$ are parametrized by the modes $\txi^a$ only, and
thus vanish if the perturbation is supersymmetric. One can check
(see appendix~\ref{appeomG}) that the equation of
motion~\eqref{eomflux} is equivalent to the equations for the modes $\txi_{5,6,7}$. Those expressions are valid for all $\tau$ and can
be evaluated by numerically integrating the explicit solution for the
$\txi^a$ and $\tphi^a$ modes found in~\cite{Bena:2011wh}.

\subsection{Infrared behavior}\label{sebsecinfrared}

We now discuss the behavior of the three--form flux in the
near--brane region, namely at small $\tau$, and we will show that both the
ISD and IASD modes are singular. The presence of
a singularity in the IASD flux mode was first noticed in~\cite{McGuirk:2009xx,Bena:2009xk}.  An explanation of
this behavior was given in~\cite{McGuirk:2009xx,Dymarsky:2011pm}, where the singularity was interpreted as coming from the coupling of
anti--D3 branes to the mode $\Phi_{-}$, which is singular in the
linearized solution, as we will show in~\eqref{phimir}. We remark that the $G_{+}$ mode also
presents a singularity at linearized level and discuss the
possible implications of this behavior. \\

The infrared expansions
for the modes $\txi^a$ and $\tphi^a$, as well as the anti--D3 boundary
conditions can be read from~\cite{Bena:2011wh}. For the IASD flux $G_{-}$ we
only need the expansions for the scalars conjugate to the flux
perturbation modes $\txi_{5,6,7}$ which are given
in~\eqref{xiinfraredexp}. From them and~\eqref{Gmfirstorder} we get
\begin{equation}\label{Gminusir}
G_{-}^1 = \frac{1}{\tau}\bigg( \frac{32}{3} \left(\frac23\right)^{1/3}P h_0
  X_1\bigg)\left( g_3\wedge g_4\wedge g_6 +3 i g_3 \wedge g_4 \wedge
    g_5\right)   + \mathcal{O}(\tau^0) \, ,
\end{equation}
where $X_1$ is defined in~\eqref{X1def}.
For the ISD flux $G_{+}$ we also need the expansions for the modes
$\tphi^a$ which can be found in section 6.4 of~\cite{Bena:2011wh}. The final result is
\begin{equation}\label{Gplusir}
G_{+}^1 = \frac{1}{\tau}\bigg( \frac{32}{3} \left(\frac23\right)^{1/3}P h_0
  X_1\bigg)\left( g_3\wedge g_4\wedge g_6 +3 i g_3 \wedge g_4 \wedge
    g_5\right) +  \mathcal{O}(\tau^0) \, .
\end{equation}
We note that $G_{+}$ shows the same kind of singularity as the $G_{-}$
mode. However we remark that, as we can see from~\eqref{Gpfirstorder},
two contributions enter in~\eqref{Gplusir}: one is from the $\txi^a$
modes, the other is from the $\tphi^a$ terms and both give rise to the
singularity. We are now going to rederive these results in a way that
will makes clear their interpretation. 

Let us introduce a set of
functions $\lambda(\tau)_A$ that parametrize the breaking of the ISD
condition~\eqref{ISDcondition}
\begin{equation}
H_3 = -\sum_A\lambda(\tau)_A \,e^{\phi} \star F_3^A \, ,
\end{equation}
where the index $A$ runs over the components of the three--forms. A
straightforward calculation shows that the ISD and IASD fluxes are
given by
\begin{equation}\label{Gpmlambda}
G_{\pm} = \sum_{A} \Big[ \big(1\pm \lambda(\tau)_A\big) \star F_3^A + i \big(\pm 1
+\lambda(\tau)_A\big) F_3^A \Big] \, .
\end{equation}
The functions $\lambda(\tau)_A$ can be obtained from the
Ansatz~\eqref{H3KS},~\eqref{F3KS}. By expanding at first--order in
$\gamma = p/M$ around the Klebanov--Strassler solution (for which the
fluxes are imaginary--self--dual), one finds the following
non--vanishing components:
\begin{align}
\lambda(\tau)^{345} &= -\frac{e^{-2y-\phi}\,f'}{2P-F} = 1 +\frac{2e^{-4A_0}}{2P-F_0}  (\txi_5+\txi_6)+\mathcal{O}(\gamma^2)\, ,\\
\lambda(\tau)^{125} & =  -\frac{e^{2y-\phi}\,k'}{F} = 1 +\frac{2e^{-4A_0}}{F_0}(\txi_5-\txi_6)+\mathcal{O}(\gamma^2)\, ,\\
\lambda(\tau)^{136}& = \lambda(\tau)^{246} =
\frac{e^{-\phi}(f-k)}{2F'} =1 + \frac{4e^{-4A_0}}{f_0-k_0} \txi_7+
\mathcal{O}(\gamma^2) \, .
\end{align}
Recall that the legs 1 and 2 are on the shrinking $S^2$, while legs 3,
4 and 5 are on the $S^3$.
While these expressions are valid for the whole conifold, we need their
near--tip behavior. The infrared expansions~\eqref{xiinfraredexp} yields
\begin{equation}\label{lambdair}
\lambda(\tau)^{345} \sim  \frac{1}{\tau}\bigg( 16 \left(\frac23\right)^{1/3}P h_0
  X_1\bigg)\, ,\quad \lambda(\tau)^{125} \sim  -\frac{1}{\tau}\bigg( 16\left(\frac23\right)^{1/3}P h_0
  X_1\bigg)  \, ,
\end{equation}
while $\lambda(\tau)^{136} = \lambda(\tau)^{246} = \mathcal{O}(\tau)$.
We thus see than only two components of $\lambda(\tau)_A$ are relevant
for the infrared physics. We can now compute $G_{\pm}^1$ by expanding
the expression~\eqref{Gpmlambda} at first--order in $\gamma$. We find
\begin{align}
G_{+}^1 &= \sum_A \Big[2 \,(\star F_3^{A})^1+ \lambda(\tau)^1_A(\star F_3^A)^0
+ i \lambda(\tau)_A^1 (F_3^A)^0\Big] \, ,\label{Gplusl}\\
G_{-}^1 &= \sum_A \Big[-\lambda(\tau)^1_A(\star F_3^A)^0
+ i \lambda(\tau)_A^1 (F_3^A)^0\Big] \, ,\label{Gminusl}
\end{align}
where we indicated by the superscript 0,1 the order of the expansion
in $\gamma$. We can now analyse the near--tip behavior of the ISD and
IASD fluxes, namely find the leading terms in an expansion near
$\tau=0$. We are interested in the origin of the singular behavior of
such modes. 

For the imaginary part, we see from the infrared expansions of the
KS fields~\eqref{irKS} that the only component that contributes to the
singularity is $F_{3}^{345}$. Since $2P - F_0 \sim 2P-\tau^2/6$, from
$\lambda(\tau)^{345}$ in~\eqref{lambdair} we recover the imaginary part
of the $G_{\pm}$ fluxes~\eqref{Gminusir},~\eqref{Gplusir}. For the
real part, we find that the only relevant component is $F_3^{125}$. We have
$(\star F_3^{125} )^0 =e^{-2y_0}\,F_0 \,g_3\wedge g_4 \wedge g_6
\sim \frac{2}{3}P \,g_3\wedge g_4 \wedge g_6$, while $(\star F_3^{125}
)^1 =e^{-2y_0} (\tphi_7 -2F_0\,\tphi_2) \,g_3\wedge g_4 \wedge
g_6$. Since 
\begin{equation}
e^{-2y_0} (\tphi_7 -2F_0\,\tphi_2)  =  \frac{1}{\tau}\bigg( \frac{32}{3} \left(\frac23\right)^{1/3}Ph_0
  X_1\bigg) + \mathcal{O}(\tau^0) \, ,
\end{equation}
we see that the two terms in the real part of $G_{+}$~\eqref{Gplusl}
give the same singularity as in the real part of $G_{-}$, in agreement
with~\eqref{Gminusir},~\eqref{Gplusir}. Before discussing the
interpretation of these results, let us show the expansions of the modes $\Phi_{\pm}$ at the first
order in $\gamma$
\begin{align}
\Phi_{-}^1 &= -\frac{1}{\tau}\bigg(16\left(\frac23\right)^{1/3}X_1\bigg) +
  \frac{32 j_0 X_1}{3} -\frac{32}{15}\left(\frac23\right)^{1/3}X_1
  \tau + \mathcal{O}(\tau^2) \, ,\label{phimir}\\
\Phi_{+}^1 & = -\frac{32j_0 X_1}{3} - \frac{4Y_4^{IR}}{h_0^2 P^4} +
\mathcal{O}(\tau^2) \, .
\end{align}
The singularity in $\Phi_{-}^1$ is expected
since the Green's function $j(\tau)$ at the linearized level diverges
at the tip. The regular behavior of $\Phi_{+}^1$ is one of the infrared
boundary conditions that was imposed in~\cite{Bena:2011wh} and ensures
that no regular D3 branes are present at the tip.
\begin{figure}[t]
\begin{center}
\includegraphics[scale=0.7]{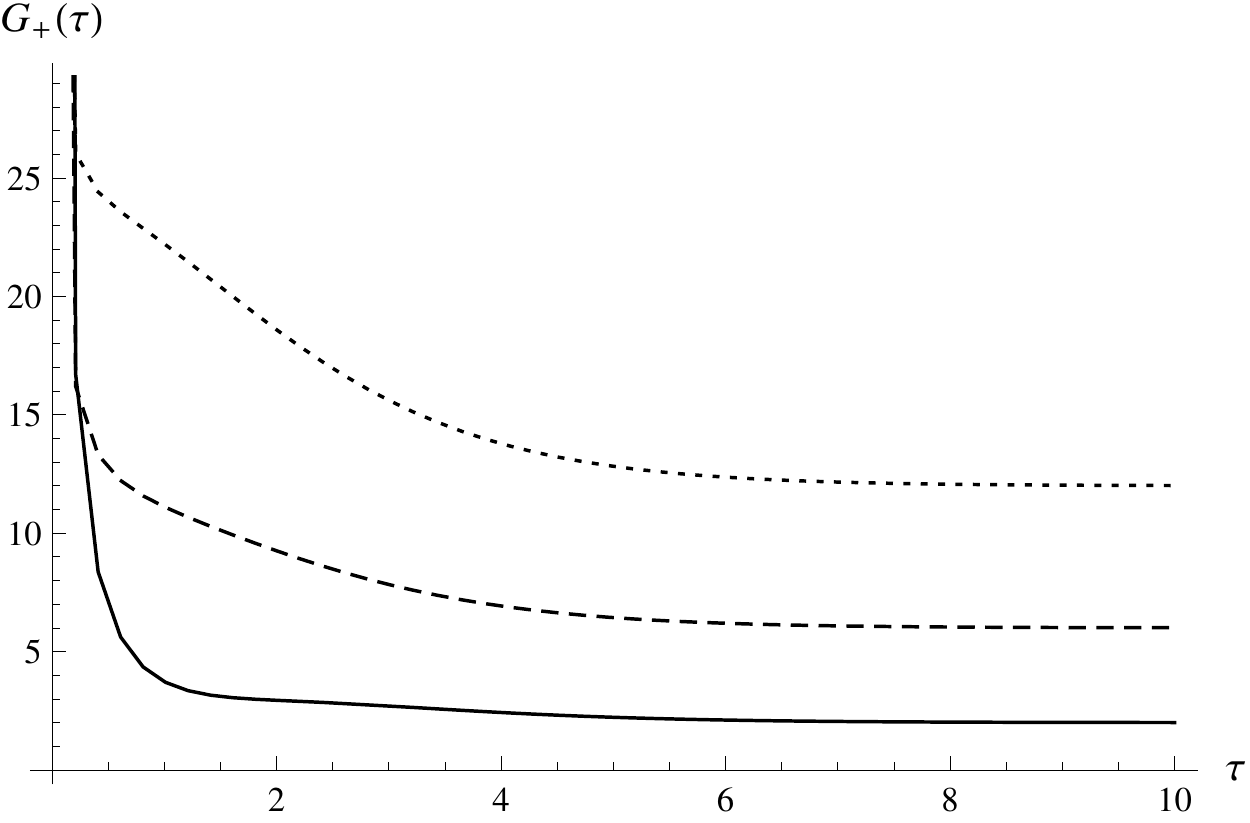} 
\caption{The ISD flux with legs on $g_3$, $g_4$ and $g_5$ for the Klebanov--Strassler geometry perturbed
  by anti--D3 branes. The plot is for $P=1,3,6$ (solid, dashed and
  dotted lines) and $p=1$.} \label{plotGplus}
\end{center}
\end{figure}

Let us summarize our findings. The ISD and IASD three--form fluxes in
the linearized anti--D3 solution have a singularity of order
$\tau^{-1}$ in the
infrared. Another mode in the solution has the same $\tau^{-1}$
singularity, namely the mode $\Phi_{-}$ which is coupled to the
anti--D3 branes. 

We can see that the singularity in $G_{-}^1$ compensates the
singularity in $\Phi_{-}$ in the equation of motion~\eqref{eomflux}~\cite{Dymarsky:2011pm,McGuirk:2009xx}. Indeed, at $\tau \sim 0 $ we find
\begin{equation}
 \Phi_{+}^0 G_{-}^1+ \Phi_{-}^1 G_{+}^0  = \mathcal{O}(\tau) \, .
\end{equation}
Based on this observation, it was argued
in~\cite{Dymarsky:2011pm,McGuirk:2009xx} that at the non--linear
level, since $\Phi_{-}$ will be finite at the tip, the $G_{-}$
singularity will disappear in the full backreacted solution.
We remark however that in the linearized solution also the $G_{+}$ mode~\eqref{Gplusir}
have a singular behavior near $\tau = 0$, as shown in
figure~\ref{plotGplus}.

 A similar situation was found for the full
backreaction of anti--D6 branes
in~\cite{Blaback:2011pn}.
While this latter setup differs in many
aspects from the Klebanov--Strassler background, it displays the same
kind of singular behavior of our linearized solution, as we will now explain. One can perform three T--dualities
along the worldvolume of the anti--D6 branes and finds that this setup
will describe anti--D3 branes on $\mathbb{R}^3 \times T^3$. If one
regards the three--torus as a large radius limit of the finite $S^3$
at the tip of the Klebanov--Strassler throat, we expect that the
anti--D6 solution will describe the behavior of the three--form flux
$F_3$ with legs on the three--sphere. From the result
of~\cite{Blaback:2011pn}, we then expect that for this flux the
\emph{full} backreacted solution will be described by the relation
\begin{equation}\label{lambdaD6}
H = -\lambda(\tau)\, e^{\phi} \star F_3 \, ,
\end{equation}
with a divergent $\lambda(\tau)$ in the near--brane region (with $\lambda(\tau) \rightarrow +\infty$). We can now
compare this expectation to our result for the linearized anti--D3
solution~\eqref{lambdair}. We see that the three--form flux with legs
on the $S^3$ (i.e. the
$g_3\wedge g_4\wedge g_5$ component) is precisely described by a relation of the
form~\eqref{lambdaD6}, with $\lambda(\tau) =
\lambda(\tau)^{345}$. As we established
in~\eqref{Gplusl},~\eqref{Gminusl}, this analogy would point towards a divergency in
the imaginary part of both the ISD and IASD fluxes at the full non--linear order. However, the leg
structure of the three--form flux in our linearized anti--D3 solution
is more complicated, and there is another component of the flux,
$F_{3}^{125}$, which contributes to the singularity in
$\lambda(\tau)$, making the anti--D6 analogy alone not
fully conclusive. 

Recently,
a discussion on the interpretation of the behavior described
by~\eqref{lambdaD6} appeared in~\cite{Blaback:2012nf}, where it was
argued that it describes an $H$--flux accumulation which 
will eventually lead to a critical value for which the barrier against brane/flux
annihilation is destroyed. It would be interesting to confirm by a
full non--linear near--brane analysis whether this
picture is valid for anti--D3 branes. While in principle the IASD flux singularity may
disappear in the full backreaction by the argument presented
in~\cite{Dymarsky:2011pm,McGuirk:2009xx}, we still find a singular ISD
flux which may survive at the non--linear order. 

 A possible way out, as schematically
depicted in~\cite{DeWolfe:2004qx,Dymarsky:2011pm}, is to argue that in
the very near--tip region the solution might be altered by the
polarization process in which the anti--branes form a fuzzy
five--brane wrapping an $S^2 \subset S^3$. In particular, one may hope
that the three--form flux singularity will be cured much as in the Polchinski--Strassler
solution~\cite{Polchinski:2000uf}. If the geometry is smoothed--out in
this way, then the effects of the backreaction will alter only
quantitatively the KPV model, by making the bound on $p/M$ for the
existence of a metastable state more strong. However, at least for smeared
anti--D3 branes, the singular ISD flux~\eqref{Gplusir} does not have
the correct legs needed for brane polarization, as also
mentioned in~\cite{Bena:2009xk}.

%%%%%%%%%%%%%%%%%%%%%%%%%%%
\section{Conclusion}
%%%%%%%%%%%%%%%%%%%%%%%%%%%

We have computed the ISD and IASD fluxes for
the linearized backreaction of $p$ anti--D3 branes on the
Klebanov--Strassler geometry. Both these modes have an infrared
singularity. While it has been suggested~\cite{Dymarsky:2011pm,McGuirk:2009xx} that
the IASD mode may be regular in the full non--linear solution, this
argument does not apply to the ISD flux. In fact, by analogy with the anti--D6
backreaction~\cite{Blaback:2010sj,Blaback:2011nz,Blaback:2011pn}, one
can even argue that some components of the
IASD flux could be singular at the non--linear level as well. It would
be interesting to verify this by computing the full backreaction near
the anti--D3 branes. If confirmed, one should
address the question whether the interpretation of this singularity is
compatible or not with the existence of a metastable state.

%%%%%%%%%%%%%%%%%%%%%%%%%%%%%%%%%%%%%%%%%%%%%
\vspace{0.5cm}
\noindent {\bf Acknowledgements}:
 \noindent 
I am grateful to Iosif Bena, Anatoly Dymarsky, Gregory Giecold,
Mariana Gra\~na, Stanislav Kuperstein and Thomas Van Riet for useful
discussions and helpful comments.
This work is supported by a Contrat de Formation par la Recherche of
CEA/Saclay and by the ERC Starting Independent Researcher Grant 259133 -- ObservableString. 
%%%%%%%%%%%%%%%%%%%%%%%%%%%%%%%%%%%%%%%%%%%%%

%%%%%%%%%%%%%%%%%%%%%%%%%%%%%%%%%%%%%%%%%%%%%%%%%%%%%%%%%%%
%%%%%%%%%%%%%%%%%%%%%%%%%%%%%%%%%%%%%%%%%%%%%%%%%%%%%%%%%%%%
\appendix

\section{Notations}\label{notation}

\subsection{Klebanov--Strassler Ansatz}
The line element in the Klebanov--Strassler Ansatz is
\begin{equation}
ds_{10}^2 = e^{2A}\, ds_{1,3}^2 + e^{-6 p-x}\, g_6^2 +
e^{x+y}\, \left( g_1^2 + g_2^2 \right) + e^{x-y}\, \left( g_3^2 +
  g_4^2 \right) + e^{-6 p-x}\, g_5^2 \, ,
\end{equation}
while the fluxes are parametrized as follows
\begin{align}
H_3 &=\frac12\left(k - f\right) \left(g_1\wedge g_3 \wedge
  g_5 + g_2 \wedge g_4 \wedge g_5\right) + f'\,g_1 \wedge g_2
\wedge g_6 + k'\, g_3 \wedge g_4 \wedge g_6 \, , \\\label{H3KS}
F_3 & = F\, g_1 \wedge g_2 \wedge g_5 + (2P-F) g_3 \wedge
g_4 \wedge g_5 + F' \left(g_1\wedge g_3 \wedge g_6 + g_2 \wedge
  g_4 \wedge g_6\right) \, , \\\label{F3KS}
F_5 & = \mathcal{F}_5 + \ast{\mathcal{F}_5}\, ,\quad\mathcal{F}_5  = \Big( F\,k + f (2P-F)\Big)
g_1\wedge g_2 \wedge g_3 \wedge g_4 \wedge g_5 \, ,\quad C_0 =0 \, , 
\end{align}
We assume that the functions $\phi^a = (x,y,p,A,f,k,F,\phi)$ only depend on the radial
variable $\tau$.
For the forms $g_i$ we use the same conventions as
in~\cite{Klebanov:2000hb}, which we reproduce here for the reader's convenience
\begin{align}
g_1 & = \frac{1}{\sqrt{2}}\Big( -\sin\theta_1 d \phi_1 -
  \cos \psi \sin \theta_2  d\phi_2+\sin \psi d\theta_2 \Big) \, ,
\\
g_2 & =  \frac{1}{\sqrt{2}}\Big( d\theta_1 - \sin\psi
  \sin\theta_2 d\phi_2 -\cos\psi d\theta_2  \Big) \, ,\nn\\
g_3 &=  \frac{1}{\sqrt{2}}\Big( -\sin\theta_1 d \phi_1 +
  \cos\psi\sin\theta_2 d\phi_2-\sin \psi d\theta_2 \Big) \, ,\non\\
g_4 &= \frac{1}{\sqrt{2}}\Big( d\theta_1 + \sin\psi
  \sin \theta_2 d\phi_2 +\cos\psi d\theta_2  \Big) \, ,\nn\\
g_5 & = d\psi + \cos \theta_2 d\phi_2 + \cos\theta_1 d\phi_1 \, , \nn\\
g_6 & = d\tau \, . \nn
\end{align}
The ISD and IASD fluxes $G_{\pm}$ and the modes $\Phi_{\pm}$ are defined
as
\begin{align}
G_{\pm} & = \star_6 G_3 \pm i G_3\, ,\qquad G_3 = F_3 + i e^{-\phi} H_3\label{G3}\\
\Phi_{\pm} &= e^{ 4A} \pm \alpha \, , \label{defPhi}
\end{align}
where
\begin{equation}
\alpha = -\int \Big[ F(\tau') k(\tau') + f(\tau')
  (2P-F(\tau'))\Big]e^{4A(\tau')-2x(\tau')} d\tau' 
\end{equation}
is the RR 4--form $C_4 = \alpha \,dx_0 \wedge \dots \wedge dx^3$ and
$e^{-4 A}$ is the warp factor\footnote{Our convention for $A$ is
  related to the one used in~\cite{Bena:2011wh} as follows:
  $2A^{\text{here}} = 2A^{\text{there}} + 2p -x$.}.
In the convention of~\cite{Bena:2011wh}, the ISD condition is
\begin{equation}\label{ISDcondition}
e^{\phi } \star F_3 + H_3 = 0 \, ,
\end{equation}
which from the definition~\eqref{G3} is equivalent to $G_{-} =
0$. 

\subsection{Linearized anti--D3 solution and infrared expansions}

The fields $\phi^a$ that enter in the KS Ansatz are expanded at
first--order in $\gamma = p/M$ around their respective background
values $\phi^a_0$
\begin{equation}
\phi^a = \phi_0^a + \phi_1^a(\gamma) + \mathcal{O}(\gamma^2) \, .
\end{equation}
The Klebanov--Strassler solution $\phi^a_0$ is given by
\begin{align} 
 e^{x_0}&= \frac14 \, h(\tau)^{1/2} \, \left( \frac12 \, \sinh(2\, \tau) - \tau \right)^{1/3} \, , \label{KSbackground}  \\
 e^{y_0}&=\tanh(\tau/2) \, , \non\\
 e^{6\, p_0}&=  24\, \frac{\left( \frac12 \, \sinh(2 \, \tau) - \tau \right)^{1/3}}{ h(\tau) \, \sinh^2\tau}  \, , \non \\
 e^{6\, A_0}&=\frac{1}{3 \cdot 2^9} \, h(\tau)\, \left( \tfrac12 \, \sinh(2 \, \tau) - \tau \right)^{2/3}\, \sinh^2\tau \, , \non\\
 f_0&= - P \, \frac{\left( \tau \, \coth \tau -1 \right)\, \left( \cosh \tau -1 \right)}{\sinh \tau} \, , \non\\
 k_0&=- P\, \frac{\left(\tau \, \coth \tau -1 \right)\, \left(\cosh \tau +1 \right)}{\sinh \tau} \, , \non \\
 F_0&= P\, \frac{\left(\sinh \tau -\tau \right)}{\sinh \tau} \, ,
 \qquad \phi_0=0 \, . \non 
 \end{align}
We define the following integrals, which correspond to the
Klebanov--Strassler warp factor $h(\tau)$ and the linearized Green's
function $j(\tau)$
\begin{align}
h(\tau) &= 32 \,P^2\,\int_\tau^{\infty} \frac{u\, \coth u - 1 }{\sinh^2 u}\, \left( \cosh u \, \sinh u-  u \right)^{1/3}\, du \, , \label{hintegral} \\
j(\tau)&
= -\int_{\tau}^{\infty} \frac{du}{\left( \cosh u \, \sinh u - u \right)^{2/3}}\, .  \label{Gintegral}
\end{align} 
We provide here the infrared expansions of the KS fields that are
relevant for computing the near--tip behavior of the three--form fluxes
\begin{align}
h(\tau) &= h_0 -\frac{16}{3}\left(\frac23\right)^{1/3}P^2 \tau^2
+ \mathcal{O}(\tau^4)\, ,\qquad e^{y_0}=\frac{\tau}{2} -\frac{\tau^3}{24}+\mathcal{O}(\tau^5) \, , \label{irKS} \\
f_0 &=-\frac{P\,\tau^3}{6}+\mathcal{O}(\tau^5) \, ,\qquad  k_0=-\frac{2\,P\,\tau}{3}-\frac{P\,\tau^3}{90}
 +\mathcal{O}(\tau^5)\, , \non\\
 F_0&= \frac{P\,\tau^2}{6} -\frac{7\,P\,\tau^4}{360}
 +\mathcal{O}(\tau^6) \, ,\non 
\end{align}
where $h_0 = h(0)\approx 18.2373\,P^2$.
For the first--order scalars, we define a rotated basis $\tphi^a$ 
\begin{equation}
 \tphi^a= \Big( -\frac32 x +3 p - 5 A,  y, x+3 p,  -2A,  f ,  k, F,
   \phi \Big) \, . \label{tphidef}
 \end{equation}
In the Hamiltonian approach, the equations of motion for the modes
$\tphi^a$ are given in terms of ``conjugate--momenta'' $\txi^a$.
We refer to~\cite{Bena:2011wh} for the definition of such modes and
for the analytic solution for the modes $\tphi^a$. Here we provide the infrared
expansions of the flux conjugate modes $\txi_5$, $\txi_6$ and $\txi_7$
which are needed in section~\ref{sebsecinfrared}. The expansions for
the $\tphi^a$ modes can be found in section 6.4 of~\cite{Bena:2011wh}.
\begin{align}
\txi_{5} & = \frac{1}{\tau}\left(8\left(\frac23\right)^{1/3} P X_1
\right) -\frac29\left(h_0+24 j_0\right) P X_1 +
  \frac{16}{15}\left(\frac23\right)^{1/3} P X_1 \tau + \mathcal{O}
  (\tau^3) \, , \label{xiinfraredexp}\\
\txi_{6} & = \frac{1}{\tau}\left(8\left(\frac23\right)^{1/3} P X_1
\right) -\frac29\left(h_0+24 j_0\right) P X_1 +
  \frac45 \, 2^{1/3} \,3^{2/3} P X_1 \tau + \mathcal{O}
  (\tau^2) \, , \non\\
\txi_7 & = -\frac{2}{27}\left(h_0-40 j_0\right) P X_1 \tau
-\frac45\,2^{1/3}\,2^{2/3} P X_1 \tau^2 +\mathcal{O}
  (\tau^3) \, ,\non
\end{align}
where $j_0 = 0.836941$.

\section{Equation of motion for $G_{\pm}$}\label{appeomG}

We want to show that the equation of motion for the fluxes
\begin{equation}
d \left(\Phi_{-}G_{+}+\Phi_{+}G_{-}\right) = 0 
\end{equation}
at the linearized level is equivalent to the equations of motion for
the supersymmetry--breaking modes $\txi_{5}$, $\txi_{6}$ and $\txi_{7}$ of~\cite{Bena:2011wh}.
The $\Phi_{\pm}$ modes that we need are
\begin{equation}
\Phi_{-}^1 = -\frac{2}{P} \txi_5 \, ,\qquad \Phi_{+}^0 = 2e^{4A_0} \, .
\end{equation}
By using the Klebanov--Strassler flow equations to eliminate derivatives of
the zeroth--order scalars, we get the following expressions in terms of the modes $\txi_{5,6,7}$
\begin{align}\label{eomGpm}
d (\Phi_{-}G_{+}&+\Phi_{+}G_{-})^1  = 4 i \,(g_1 \wedge g_2 \wedge g_5
\wedge g_6 +g_3 \wedge g_4 \wedge g_5 \wedge g_6)\\
&\times\Big[P\,\txi_7 + (-P+F_0)\,\txi_5' + P \,\txi_6' \Big]+ (g_1 \wedge g_3\wedge g_5\wedge g_6\
  +g_2\wedge g_4 \wedge g_5\wedge g_6)\non\\
&\times\Big[
\frac{2}{P}(k_0-f_0)\,\txi_5' +4\,\big(\sinh(2\, y_0)\, \txi_5 + \cosh(2\,
y_0)\, \txi_6+\txi_7'\big)\Big]\, . \non
\end{align}
If we substitute the derivatives of $\txi_a$ with their equations of motion,
which are~\cite{Bena:2011wh}
\begin{align}
\txi_5'&=-\frac13 \, P\, e^{-2\, x_0}\, \txi_1 \label{xi5eom}\\
\txi_6'&=-\txi_7-\frac13 \, e^{-2\, x_0} \, \left( P - F_0 \right)\, \txi_1 \\
\txi_7'&=-\sinh(2\, y_0)\, \txi_5 - \cosh(2\, y_0)\, \txi_6 + \frac16
\, e^{-2\, x_0}\, \left( f_0-k_0 \right)\, \txi_1 \, ,
\end{align}
we check that all the components of~\eqref{eomGpm} vanish.

%%%%%%%%%%%%%%%%%%%%%%%%%%%%%%%%%%%%%%%%%%%%%%%%%%%%%%%

\providecommand{\href}[2]{#2}\begingroup\raggedright\endgroup


\begin{thebibliography}{10}


\bibitem{Kachru:2002gs}
S.~Kachru, J.~Pearson, and H.~L. Verlinde, ``{Brane/flux Annihilation and the
  String Dual of a Non-Supersymmetric Field Theory},'' {\em JHEP} {\bf 06}
  (2002) 021,
\href{http://arXiv.org/abs/hep-th/0112197}{[hep-th/0112197]}.
%%CITATION = HEP-TH 0112197;%%.

\bibitem{Klebanov:2000hb}
I.~R. Klebanov and M.~J. Strassler, ``Supergravity and a confining gauge
  theory: Duality cascades and chisb-resolution of naked singularities,'' {\em
  JHEP} {\bf 08} (2000) 052,
\href{http://arXiv.org/abs/hep-th/0007191}{ [hep-th/0007191]}.
%%CITATION = HEP-TH 0007191;%%.

%\cite{DeWolfe:2004qx}
\bibitem{DeWolfe:2004qx}
  O.~DeWolfe, S.~Kachru and H.~L.~Verlinde,
  ``The Giant inflaton,''
  JHEP {\bf 0405} (2004) 017
\href{http://arXiv.org/abs/hep-th/0403123}{[hep-th/0403123]}.
  %%CITATION = HEP-TH/0403123;%%

%\cite{DeWolfe:2008zy}
\bibitem{DeWolfe:2008zy}
  O.~DeWolfe, S.~Kachru and M.~Mulligan,
  ``A Gravity Dual of Metastable Dynamical Supersymmetry Breaking,''
  Phys.\ Rev.\ D {\bf 77} (2008) 065011
 \href{http://arXiv.org/abs/0912.3519}{[arXiv:0801.1520 [hep-th]]}.
  %%CITATION = ARXIV:0801.1520;%%

%\cite{McGuirk:2009xx}
\bibitem{McGuirk:2009xx}
  P.~McGuirk, G.~Shiu and Y.~Sumitomo,
  ``Non-supersymmetric infrared perturbations to the warped deformed conifold,''
  Nucl.\ Phys.\ B {\bf 842} (2011) 383
 \href{http://arXiv.org/abs/0910.4581 }{[arXiv:0910.4581 [hep-th]]}.
  %%CITATION = ARXIV:0910.4581;%%

\bibitem{Bena:2009xk}
I.~Bena, M.~Grana, and N.~Halmagyi, ``{On the Existence of Meta-stable Vacua in
  Klebanov-Strassler},'' {\em JHEP} {\bf 1009} (2010) 087,
  \href{http://arXiv.org/abs/0912.3519}{[arXiv:0912.3519 [hep-th]]}.

\bibitem{Bena:2011hz}
I.~Bena, G.~Giecold, M.~Grana, N.~Halmagyi, and S.~Massai, ``{On Metastable
  Vacua and the Warped Deformed Conifold: Analytic Results},''
  \href{http://arXiv.org/abs/1102.2403}{[arXiv:1102.2403 [hep-th]]}.

%\cite{Bena:2011wh}
\bibitem{Bena:2011wh}
  I.~Bena, G.~Giecold, M.~Grana, N.~Halmagyi and S.~Massai,
  ``The backreaction of anti-D3 branes on the Klebanov-Strassler
  geometry,''
 \href{http://arXiv.org/abs/1106.6165}{arXiv:1106.6165 [hep-th]}.
  %%CITATION = ARXIV:1106.6165;%%

\bibitem{Dymarsky:2011pm}
A.~Dymarsky, ``{On gravity dual of a metastable vacuum in Klebanov-Strassler
  theory},'' {\em JHEP} {\bf 1105} (2011) 053,
  \href{http://arXiv.org/abs/1102.1734}{[arXiv:1102.1734 [hep-th]]}.


%\cite{Klebanov:2010qs}
\bibitem{Klebanov:2010qs}
  I.~R.~Klebanov and S.~S.~Pufu,
  ``M-Branes and Metastable States,''
  JHEP {\bf 1108} (2011) 035
\href{http://arXiv.org/abs/1006.3587}{[arXiv:1006.3587 [hep-th]]}.
  %%CITATION = ARXIV:1006.3587;%%

%\cite{Bena:2010gs}
\bibitem{Bena:2010gs}
  I.~Bena, G.~Giecold and N.~Halmagyi,
  ``The Backreaction of Anti-M2 Branes on a Warped Stenzel Space,''
  JHEP {\bf 1104} (2011) 120
\href{http://arXiv.org/abs/1011.2195}{[arXiv:1011.2195 [hep-th]]}.
  %%CITATION = ARXIV:1011.2195;%%

%\cite{Massai:2011vi}
\bibitem{Massai:2011vi}
  S.~Massai,
  ``Metastable Vacua and the Backreacted Stenzel Geometry,''
\href{http://arXiv.org/abs/1110.2513}{arXiv:1110.2513 [hep-th]}.
  %%CITATION = ARXIV:1110.2513;%%

%\cite{Giecold:2011gw}
\bibitem{Giecold:2011gw} 
  G.~Giecold, E.~Goi and F.~Orsi,
  ``Assessing a candidate IIA dual to metastable
  supersymmetry-breaking,''
{\em JHEP} {\bf 1202} (2012) 019,
\href{http://arXiv.org/abs/1108.1789}{arXiv:1108.1789 [hep-th]}.
  %%CITATION = ARXIV:1108.1789;%%


\bibitem{Blaback:2010sj}
J.~Blaback, U.~H. Danielsson, D.~Junghans, T.~Van~Riet, T.~Wrase, {\em et al.},
  ``{Smeared versus localised sources in flux compactifications},'' {\em JHEP}
  {\bf 1012} (2010) 043, \href{http://arXiv.org/abs/1009.1877}{arXiv:1009.1877 [hep-th]}.

\bibitem{Blaback:2011nz}
J.~Blaback, U.~H. Danielsson, D.~Junghans, T.~Van~Riet, T.~Wrase, {\em et al.},
  ``{The problematic backreaction of SUSY-breaking branes},'' {\em JHEP} {\bf 1108} (2011) 105,
  \href{http://arXiv.org/abs/1105.4879}{arXiv:1105.4879 [hep-th]}.

%\cite{Blaback:2011pn}
\bibitem{Blaback:2011pn}
  J.~Blaback, U.~H.~Danielsson, D.~Junghans, T.~Van Riet, T.~Wrase and M.~Zagermann,
  ``(Anti-)Brane backreaction beyond perturbation theory,'' JHEP {\bf 1202} (2012) 025,
\href{http://arXiv.org/abs/1111.2605}{arXiv:1111.2605 [hep-th]}.
  %%CITATION = ARXIV:1111.2605;%%

%\cite{Blaback:2012nf}
\bibitem{Blaback:2012nf}
  J.~Blaback, U.~H.~Danielsson and T.~Van Riet,
  ``Resolving anti-brane singularities through time-dependence,''
\href{http://arXiv.org/abs/1202.1132}{arXiv:1202.1132 [hep-th]}.
  %%CITATION = ARXIV:1202.1132;%%

%\cite{Polchinski:2000uf}
\bibitem{Polchinski:2000uf}
  J.~Polchinski and M.~J.~Strassler,
  ``The String dual of a confining four-dimensional gauge theory,''
\href{http://arXiv.org/abs/0003136}{[hep-th/0003136]}.
  %%CITATION = HEP-TH/0003136;%%

%\cite{Myers:1999ps}
\bibitem{Myers:1999ps}
  R.~C.~Myers,
  ``Dielectric branes,''
  JHEP {\bf 9912} (1999) 022
\href{http://arXiv.org/hep-th/9910053}{[hep-th/9910053]}.
  %%CITATION = HEP-TH/9910053;%%


\bibitem{Kachru:2003sx}
S.~Kachru {\em et al.}, ``{Towards Inflation in String Theory},'' {\em JCAP}
  {\bf 0310} (2003) 013,
\href{http://arXiv.org/abs/hep-th/0308055}{[hep-th/0308055]}.
%%CITATION = HEP-TH 0308055;%%.

%\cite{Borokhov:2002fm}
\bibitem{Borokhov:2002fm}
  V.~Borokhov and S.~S.~Gubser,
  ``Nonsupersymmetric deformations of the dual of a confining gauge theory,''
  JHEP {\bf 0305} (2003) 034
\href{http://arXiv.org/abs/hep-th/0206098}{[hep-th/0206098]}.
  %%CITATION = HEP-TH/0206098;%%

%\cite{Giddings:2001yu}
\bibitem{Giddings:2001yu}
  S.~B.~Giddings, S.~Kachru and J.~Polchinski,
  ``Hierarchies from fluxes in string compactifications,''
  Phys.\ Rev.\ D {\bf 66} (2002) 106006
\href{http://arXiv.org/abs/hep-th/0105097}{[hep-th/0105097]}.
  %%CITATION = HEP-TH/0105097;%%

\bibitem{Kuperstein:2003yt}
S.~Kuperstein and J.~Sonnenschein, ``{Analytic non-supersymmetric background
  dual of a confining gauge theory and the corresponding plane wave theory of
  hadrons},'' {\em JHEP} {\bf 02} (2004) 015,
\href{http://arXiv.org/abs/hep-th/0309011}{[hep-th/0309011]}.
%%CITATION = HEP-TH/0309011;%%.

%\cite{Halmagyi:2011yd}
\bibitem{Halmagyi:2011yd}
  N.~Halmagyi, J.~T.~Liu and P.~Szepietowski,
  ``{On N = 2 Truncations of IIB on $T^{1,1}$},''
\href{http://arXiv.org/abs/1111.6567}{ arXiv:1111.6567 [hep-th]}.
  %%CITATION = ARXIV:1111.6567;%%

\end{thebibliography}
\end{document}